\documentclass[preprint,showpacs,preprintnumbers,amsmath,amssymb,nofootinbib]{revtex4}
\usepackage{epsfig}

\newcommand{\h}{\hat{H}}
\newcommand{\U}{\hat{U}}

\newcommand*{\ket}[1]{\left| #1 \right\rangle}
\newcommand*{\bra}[1]{\left\langle{#1}\right|}

\newcommand{\A}{\mathbf{A}}
\newcommand{\K}{\mathbf{k}}

\newcommand{\R}{\mathbf{r}}

\newcommand{\kk}{{k_{\parallel}}}
\renewcommand{\k}{{k_{\perp}}}

\renewcommand{\Re}{{\rm Re}\,}
\renewcommand{\Im}{{\rm Im}\,}

\begin{document}

\title{Two-electron ionization in strong laser fields below intensity threshold: signatures of attosecond timing
in correlated spectra}

\author{Denys I. Bondar}
\email{dbondar@sciborg.uwaterloo.ca}
\address{Department of Physics and Astronomy, University of Waterloo,
200 University Avenue West, Waterloo, Ontario N2L 3G1, Canada}
\address{National Research Council, 100 Sussex Drive, Ottawa, Ontario
K1A 0R6, Canada}

\author{Wing-Ki Liu}
\email{wkliu@sciborg.uwaterloo.ca}
\address{Department of Physics and Astronomy, University of Waterloo,
200 University Avenue West, Waterloo, Ontario N2L 3G1, Canada}

\author{Misha Yu. Ivanov}
\email{Misha.Ivanov@nrc-cnrc.gc.ca}
\address{National Research Council, 100 Sussex Drive, Ottawa, Ontario
K1A 0R6, Canada}

\date{\today}

\begin{abstract}
We develop an analytical model of correlated two-electron
ionization in strong infrared laser fields. The model includes all
relevant interactions between the electrons, the laser field, and
the ionic core nonperturbatively. We focus on the deeply quantum
regime, where the energy of the active electron driven by the
laser field is insufficient to collisionally ionize the parent
ion, and the assistance of the laser field is required to create a
doubly charged ion. In this regime, the electron-electron and the
electron-ion interactions leave distinct footprints in the
correlated two-electron spectra, recording the mutual dynamics of
the escaping electrons.
\end{abstract}

\maketitle

\section{Introduction}

In strong infrared laser fields, following one-electron ionization
of an atom or a molecule, the liberated electron can recollide
with the parent ion \cite{Corkum_1993, Kuchiev_1987}. The electron
acts as an ``atomic antenna'' \cite{Kuchiev_1987}, absorbing the
energy from the laser field between ionization and recollision and
depositing it into the parent ion. Inelastic scattering on the
parent ion results in further collisional excitation and/or
ionization. Liberation of the second electron during the
recollision -- the laser-induced e-2e process -- is known as
correlated, or nonsequential, double ionization (NSDI). 

The phenomenon of NSDI was experimentally discovered by Suran and Zapesochny \cite{Suran1975} for alkaline-earth atoms
(for further experimental investigations of NSDI for alkaline-earth atoms, see, e.g., Refs.  \cite{Bondar1993, Bondar1998, Bondar2000}). In this case, autoionizing double excitations below the second ionization were shown to be extremely important. For a theoretical study of these effects, see, e.g., Ref. \cite{Lambropoulos1988}. For noble gas atoms, nonsequential double ionization was first observed by L'Huillier {\it et al.} (see, e.g., Refs. \cite{LHuillier1982, LHuillier1983}). The interest to the phenomenon of NSDI grew rapidly after NSDI was rediscovered in 1993-1994 \cite{Walker1993, Walker1994}, and
now for IR fields and higher intensities. Recently,
correlated multiple ionization has also been observed
\cite{Rudenko_2004, Zrost_2006}. The renewed interest in NSDI has been enhanced by the availability of new experimental techniques that allow one to perform accurate measurements of the angle- and energy-resolved spectra of the photoelectrons, in coincidence. Such measurements play a crucial role in elucidating the physical mechanisms behind the NSDI. 

From the theoretical perspective, direct {\it ab initio} simulations of
the photoelectron spectra corresponding to correlated (or
nonsequential) double ionization in intense low-frequency laser
fields represent a major challenge. Only now such benchmark
simulations have become possible \cite{Taylor2007} for
the typical experimental conditions (the helium atom, laser
intensity $I\sim 10^{15}$ W/cm$^2$, laser wavelength $\lambda =
800$ nm).

What are the physical reasons behind these numerical challenges,
which push the modern computational resources to their limit,
occupying thousands of processors for weeks at a time? First, they lie in the need to deal with highly nonstationary
two-electron dynamics, with electron energies changing by hundreds
of eV on a subfemtosecond time scale, the characteristic
amplitudes of electron oscillations reaching several tens of
angstroms, and final electron energies ranging from zero to $10^3$
eV. Accurate description of such dynamics requires
attosecond-scale time steps, very large grids, and small grid steps
$\sim 0.1$ $\AA$.

Second, one needs to analyze the results of such massive
calculations, which output a five-dimensional, time-dependent
wave function (one spatial dimension is saved by the cylindrical
symmetry of the problem in a linearly polarized laser field.)
Extracting essential physical processes and mechanisms responsible
for correlated double ionization from such massive data arrays is
a separate and equally formidable challenge.

In addition to the experimental measurements, especially of the
correlated electron spectra \cite{Staudte_2007, Rudenko_2007, Liu_2008, Zrost_2006, Zeidler_2005, Weckenbrock_2004, Rudenko_2004},  tremendous insight into the physics of the problem
has been obtained from classical simulations performed in Refs.
\cite{Panfili_2002, Phay2005, Ho_2005, Haan2006, Haan2008, Ho2006}. These papers have demonstrated a  variety of the
regimes of nonsequential double and triple ionization. Not only
do these simulations reproduce key features observed in the
experiment, they also give a clear view of the (classical)
interplay between the two electrons, the potentials of the laser
field, and of the ionic core. They also show how different types of
the correlated motion of the two electrons contribute to different
parts of the correlated two-electron spectra.

Our goal is to develop a fully quantum, analytical treatment of
this problem. It is important that the approach takes into account
all relevant interactions -- those with the laser field, the ion,
and between the electrons -- nonperturbatively. Given the
complexity of the problem, it is clear that the  analytical
description will have to incorporate physical understanding of the
dynamics gained from the previous experimental and theoretical
work.

In particular, the physics of double ionization is different for
different intensity regimes, separated by the ratio of the energy
of the recolliding electron to the binding (or excitation) energy
of the second electron, bound in the ion.

The maximum energy, which the recolliding electron can acquire
from the laser field, is $\sim 3.2 U_p$ \cite{Corkum_1993}, where
$U_p=F^2/4\omega^2$, $F$ is the laser field strength, and $\omega$
is the laser frequency (atomic units are used throughout the
paper). Even when $3.2 U_p$ is far from sufficient to liberate
other electrons, experiments have observed correlated ionization
\cite{Zeidler_2005, Weckenbrock_2004, Rudenko_2004, Zrost_2006,
Liu_2008}.  As opposed to the more conventional high-$U_p$ regime
(see, e.g., Refs. \cite{Staudte_2007, Rudenko_2007, Becker_2005,
Yudin_2001A, Lein2000,deMorissonFaria2008a, deMorissonFaria2008b,
deMorissonFaria2005, deMorissonFaria2004, deMorissonFaria2004B,
deMorissonFaria2003, Liu_2006} and references therein), in the
low-$U_p$ regime the assistance of the laser field during the
recollision is crucial.

Here, we focus on this most challenging low-$U_p$ regime, where
the nonperturbative interplay of all three interactions is
crucial. In this regime, existing classical and quantum analysis
(see, e.g., Refs. \cite{Haan2006, Ho2006, deMorissonFaria2006}) demonstrates two possibilities of electron ejection
after the recollision. First, the two electrons can be ejected
with little time delay compared to the quarter-cycle of the
driving field. Second, the time delay between the ejection of the
first and the second electron can approach or exceed the
quarter-cycle of the driving field. In these two cases, the
electrons appear in different quadrants of the correlated
spectrum. If, following the recollision, the electrons are ejected
nearly simultaneously, their parallel momenta have equal signs,
and both electrons are driven by the laser field in the same
direction toward the detector. If, following the recollision, the
electrons are ejected with a substantial delay (quarter-cycle or
more), they end up going in the opposite directions. Thus, these
two types of dynamics leave distinctly different traces in the
correlated spectra.

Here, we consider the case in which the two electrons are ejected
nearly simultaneously. We show that in this case the correlated
spectra should bear clear signatures of the electron-electron and
electron-ion interactions after ionization,  including the
interplay of these interactions. We identify these signatures.

To study NSDI analytically, we use the strong-field eikonal-Volkov
approach (SF-EVA) and follow the recipe described in  Ref.
\cite{Smirnova_2008}.  Our model complements earlier theoretical
work  on calculating correlated two-electron spectra using
the strong-field S-matrix approach \cite{Becker_2005}. The key
theoretical advance of this work is the ability to include
nonperturbatively all relevant interactions for both active
electrons: with each other, with the ion, and with the laser
field. Electron-electron and electron-ion interactions are
included on an equal footing. Our model ignores multiple
recollisions and multiple excitations developing over several
laser cycles, such as those seen in the classical simulations
\cite{Ho_2005}. This simplification is particularly adequate for
the few-cycle laser pulses, as demonstrated in the experiment
\cite{Bhardwaj_2001}. According to this experiment, multiple
recollisions are noticeably suppressed already for 12 fs pulses
at $\lambda=800$ nm. For 6--7 fsec pulses at $\lambda=800$ nm, such
simplification is justified.

The rest of this paper is organized as follows: In Sec. \ref{s0},
preliminarily comments from the point of view of the Landau-Dykhne
adiabatic theory are made. Basing on the SF-EVA, we develop the
analytical model of NSDI in Sec. \ref{s1}. The
correlated spectra are first calculated within the strong field
approximation (SFA) in Sec. \ref{s3}. In Secs.
\ref{s4} and \ref{s5}, the roles of electron-electron and
electron-ion interactions are analyzed. We show how these
interactions lead to profound qualitative differences from the
SFA-based models in positions and shapes of maxima in the
correlated spectra. Finally, the conclusions are made in the final
section.

\section{Fundamental Principles and Background}\label{s0}

\begin{figure}
\begin{center}
\includegraphics[scale=0.3]{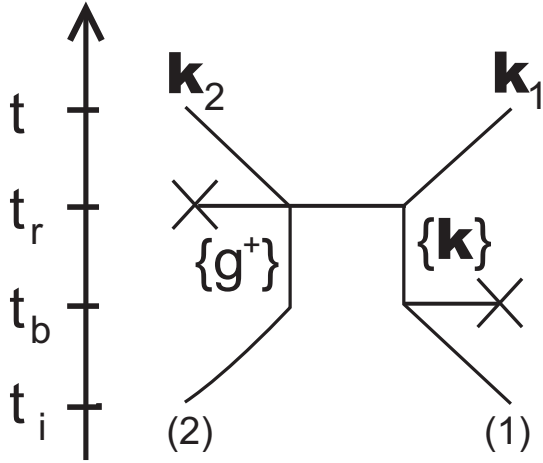}
\end{center}
\caption{The diagram of NSDI within the considered regime.}
\label{Fig4}
\end{figure}

The process is shown by the Feynman diagram in Fig. \ref{Fig4}. The
system begins in the ground state $\ket{gg}$ at time $t_i$. At an
instant $t_b$, intense laser field promotes the first electron to
the continuum state $\ket{\K}$; the second electron remains in the
ground state of the ion $\ket{g^+}$. Recollison at $t_r$ frees
both electrons. The symmetric diagram where electrons 1 and 2
change their roles is not shown, but is included in the calculated
spectrum.

Despite the fact that the main approach used in this paper is the
strong-field eikonal Volkov approximation (SF-EVA), it is
methodologically useful to look first at the problem at hand from
the point of view of the Landau-Dykhne (LD) adiabatic
approximation. Indeed, since the frequency of the laser field is
 low compared to other relevant energy scales such as the binding energy,
the LD approximation is a natural way of tackling the problem.
Moreover, the LD approach is an especially handy tool for
obtaining results within the strong field approximation, which is
used as the zero-order approximation in the SF-EVA
\cite{Smirnova_2008}.

According to the LD method \cite{Dykhne_1962,
Landau_1977} (see also Ref. \cite{Delone_1985}), if the
Hamiltonian of a system $\h(t)$ is a slowly varying function of
time $t$, and $\h(t) \psi_n(t) = E_n(t) \psi_n(t)$ $(n=i,f)$,
then the probability $\Gamma$ of the transition $\psi_i \to
\psi_f$ is given by (the atomic units are used throughout)
\begin{equation}\label{DykhneApp}
\Gamma \propto \exp\left( -2 \Im \int_{t_1}^{t_0} \Big[E_f(t) -E_i(t)\Big] dt \right),
\end{equation}
where $t_1$ is any point on the real axis of $t$, and $t_0$ is the complex root of the equation
\begin{equation}\label{EqT0}
E_i(t_0) = E_f(t_0),
\end{equation}
which lies in the upper half-plane. If there are several roots, we
must choose the one that is the closest to the real axis of $t$.
It must be stressed that there are no assumptions regarding the
form of the Hamiltonian. Further discussions and generalizations
of the LD method can be found in Ref. \cite{Chaplik_1964,
Davis_1976, Joye_1993, Moyer_2001, Schilling_2006}.

The LD approach has many applications in different areas of
physics. In particular, it has been extensively used in strong
field physics \cite{Delone_1985}. For example, the problem of
single-electron ionization can be analyzed within the LD
approximation (see Ref. \cite{Bondar2008, Rastunkov_2007, Krainov_2003, Yudin_2001B,  Delone_1991, Delone_1985} and references therein) by
setting $E_i = -I_p$ to be the energy of the ground state ($I_p$
is the ionization potential) and $E_f(t) = \left[ \K + \A(t)
\right]^2/2$ to be the energy of the free electron, oscillating in
the laser field.  Here $\K$ is the final momentum of the electron
at the detector, and
$$
 \A(t) = - ({\bf F}/\omega)\sin(\omega t)
$$
is the vector potential. The results of the LD approach in this
form are fully consistent with the usual SFA. Improving the SFA
result by incorporating the Coulomb potential is also possible, as
we discuss below for the case of two liberated electrons.

Let us apply the general approach Eq. (\ref{DykhneApp}) to the
two-electron process under consideration. The NSDI has two stages, namely
ionization of the first electron and the recollision. Hence,
strictly speaking, the LD adiabatic approximation has to be
applied to each of the two stages, since the total amplitude of
the process is the product of the ionization amplitude and the
recollision amplitude. However, it is the second (recollisioin)
amplitude that is responsible for the shape of the correlated spectra.
The first amplitude only gives the overall height of the spectra,
as it determines the overall probability of the recollision. Since
at this stage we are only interested in the shape of the
correlated spectra, we omit the ionization amplitude from this
discussion (it is included later in the full treatment).

As a zero approximation, we define $E_{i}(t)$ and $E_{f}(t)$ for
the second part of NSDI without the Coulomb interaction. Before
the recollision at the moment $t_r$ (Fig. \ref{Fig4}), one electron
is bound and another is free. The classical energy of the system
before $t_r$ is
\begin{equation}\label{EiCMI}
E_i (t)= \frac 12 \left[ -\A(t_b(t))+\A(t) \right]^2 + E_{g^+},
\end{equation}
where $E_{g^+}$ denotes the energy level of the second (bound)
electron. The time of ``birth'' (ionization) for the first electron
$t_b(t)$ is the standard function of the instant of recollision $t_r$,
which is obtained from the saddle-point S-matrix calculations in the Appendix. In Eq. (\ref{EiCMI}), we have assumed that
the recolliding electron has been born at $t_b(t_r)$ with zero
velocity. After the recollision, both electrons are free and the
energy of the system is
\begin{equation}\label{EfCMI_}
E_f (t) = \frac 12  [\K_1 + \A(t)]^2 +  \frac 12  [\K_2 + \A(t)]^2,
\end{equation}
where $\K_{1,2}$ are the asymptotic kinetic momenta at
$t\to\infty$ of  the first and second electrons, correspondingly.

Now, substituting Eqs. (\ref{EiCMI}) and (\ref{EfCMI_}) into Eq.
(\ref{DykhneApp}), we obtain the correlated spectrum standard for
the strong field approximation (SFA),
\begin{eqnarray}\label{SFASpectra}
\Gamma_{SFA}(\K_1, \K_2) &\approx& \exp\left(-\frac 2{\omega}\Im S_{SFA}(\K_1, \K_2)\right), \\
S_{SFA}(\K_1, \K_2)&=&  \int_{\Re\varphi_r^0}^{\varphi_r^0}
\left[  \frac 12 \left(\K_1 + \A(\varphi)\right)^2 +  \frac 12 \left(\K_2 + \A(\varphi)\right)^2
  -\frac 12 \left[ \A(\varphi)-\A(\Phi(\gamma; \varphi)) \right]^2 + I_p^{(2)} \right] d\varphi,
\nonumber
\end{eqnarray}
where the phase of ``birth'' (ionization) $\Phi(\gamma; \varphi)$ corresponding to the recollison phase $\varphi$ and the transition point $\varphi_r^0$ are defined by Eqs. (\ref{QuantPhaseBirth}) and (\ref{DeltaEGamma}) in the Appendix. We will come back to more rigorous analysis of the same spectra in the next section. 

The major stumbling block is to
account for the electron-electron and the electron-ion
interactions on the same footing, nonperturbatively. To include
these crucial corrections, we have to include the corresponding
Coulomb interactions into $E_{i,f}(t)$. With the nucleus located
at the origin, the electron-electron and the electron-core
interaction energies are
\begin{equation}\label{PotentialEnergies}
V_{ee} = 1/|\R_{12}(t)|, \qquad V_{en}^{(1,2)} = -2/|\R_{1,2}(t)|,
\end{equation}
correspondingly. Here $\R_{12}(t) = \R_1(t) - \R_2(t)$ and
$\R_{1,2}(t)$ are the trajectories of the two electrons.

However, we immediately see problems. The corrections depend on
the specific trajectory, and one needs to somehow decide what
this trajectory should be. Note that the classical trajectories
$\R_{1,2}(t)$ in the presence of the laser field and the Coulomb
field of the nucleus may even be chaotic. The solution to this
problem has already been discussed in the original papers by Popov
and co-workers  \cite{Perelomov_1966, Perelomov_1967_A, Perelomov_1967_B,
Perelomov_1968} for single-electron ionization. In the
spirit of the eikonal approximation, these trajectories can be
taken in the laser field only \cite{Smirnova_2008, Perelomov_1968, Perelomov_1967_B}, so that they correspond to the saddle points of the
standard SFA analysis. Not surprisingly, in the SFA these
trajectories start at the origin,
$$
\R_{1,2}(t) = \int_{t_0}^{t} [ \K_{1,2} + \A(\tau)] d\tau.
$$

However, here we run into the second problem: the potentials
$V_{ee}$ and $V_{en}^{(1,2)}$ are singular. Consequently,  the
integral in Eq. (\ref{DykhneApp}) is divergent and the result is
unphysical. Therefore, such implementation of the Coulomb
corrections requires additional care.

The next sections describe a rigorous approach that deals with
these two problems, both defining the relevant trajectories and
removing the divergences of the integrals.

\section{Basic Formalism}\label{s1}

The key step in dealing with the singularities of the Coulomb
potentials during the recollision is to partition the
electron-electron and electron-ion interactions in the
two-electron Hamiltonian as follows:
\begin{eqnarray}\label{PotentialsIntro}
V_{ee} &\equiv& V_{ee} - V_{ee,lng} + V_{ee,lng} = V_{ee,lng} + \Delta V_{ee,shr}, \nonumber\\
V_{en} &\equiv& V_{en} - V_{en,lng} + V_{en,lng} = V_{en,lng} + \Delta V_{en, shr}.
\end{eqnarray}
The potential $V_{ee,lng}$ has a long-range behavior identical to $V_{ee}$, but no
singularity at the origin, and $\Delta V_{ee, shr}$ is singular but short-range potential.
The same applies to $V_{en,lng}$ and $\Delta V_{en,shr}$. We choose
\begin{equation}\label{Potentials}
\Delta V_{en,shr}(r) = V_{en}(r)\exp(-r/r_0), \quad \Delta
V_{ee,shr}(r_{12}) = V_{ee}(r_{12})\exp\left[ -
r_{12}/r_{12}^{(0)}\right],
\end{equation}
where $r_0$ and $r_{12}^{(0)}$ will be defined later.  Note that
the partitioning (\ref{PotentialsIntro}) and (\ref{Potentials}) has
been employed originally in the Perelomov-Popov-Terent'ev approach
 \cite{Perelomov_1966, Perelomov_1967_A, Perelomov_1967_B,
Perelomov_1968} for the problem of single-electron ionization.

Now, we can write the Hamiltonian as
$$
\h(t) = \h_s(t) + \Delta V_{shr},
$$
where $\Delta V_{shr} \equiv \Delta V_{ee,shr} + \Delta V_{en, shr}$ and $\h_s$ is the rest,
which includes smoothed Coulomb potentials for
electron-electron and electron-nuclear interactions, $V_{ee,lng}$ and $V_{en,lng}$.

To first order in $\Delta V_{shr}$, the amplitude to find two electrons with momenta $\K_1, \K_2$
at the detector at the time $t$ is
\begin{eqnarray}\label{Ampl}
a(\K_1, \K_2) = -i\int_{t_i}^t dt_r \int d^3 \K \bra{ \K_1 \K_2} \U_s (t, t_r)\Delta V_{shr} \ket{g^+ \K}
\bra{\K g^+}\U_s (t_r, t_i) \ket{gg}.
\end{eqnarray}
Approximations in Eq. (\ref{Ampl}) are first order in $\Delta V_{shr}$ and the assumption that
at the moment of recollision the ion is in its ground state. Both are well justified.

The next step is to approximate the two parts of the evolution:
before $t_r$ and after $t_r$. The key component for correlated
spectra is the second part -- after $t_r$. The main aspect of the
first part of the evolution -- prior to $t_r$ -- is to supply
an active electron with the required energy.

To simplify the amplitude $\bra{ \K_1 \K_2} \U_s (t, t_r)\Delta V_{shr} \ket{g^+ \K}$,
we insert the decomposition of unity,
\begin{eqnarray}\label{AmplB}
&& b(\K_1, \K_2, \K, t_r) = \bra{ \K_1 \K_2} \U_s (t, t_r)\Delta V_{shr} \ket{g^+ \K} = \nonumber\\
&&= \int\int d^3 \R_1 d^3 \R_2 \bra{\K_1 \K_2} \U_s(t,t_r) \ket{\R_1 \R_2}\bra{\R_1 \R_2} \Delta V_{shr} \ket{g^+ \K}\approx \nonumber\\
&&\approx  \int\int d^3 \R_1 d^3 \R_2  \bra{\K_1 + \A(t_r) ,\, \K_2 + \A(t_r)} \R_1 \R_2 \rangle \bra{\R_1 \R_2} \Delta V_{shr} \ket{g^+ \K}
\times \nonumber\\
&&\exp\left[ -i\int_{t_r}^t \left\{ \frac 12 [\K_1 + \A(\tau)]^2 + \frac 12 [\K_2 + \A(\tau)]^2 +\right.\right. \nonumber\\
&& \qquad +V_{ee,lng}(\R_{12}(\tau))+ V_{en,lng}(\R_{1}(\tau))+V_{en,lng}(\R_{2}(\tau)) \Big\}d\tau\Big].
\end{eqnarray}
Here we have applied the SF-EVA method \cite{Smirnova_2008}.  The
integral from the nonsingular parts of the electron-electron and
electron-ion interactions are calculated along the trajectories in
the laser field only. The trajectories
\begin{equation}\label{Trajectories}
\R_{1,2}(t) = \R_{1,2} + \int_{t_r}^{t} [ \K_{1,2} + \A(\tau)] d\tau
\end{equation}
and $\R_{12}(t) = \R_1(t)-\R_2(t)$ begin at the positions $\R_1,
\R_2$ at instant $t_r$.  The bra-vectors $\bra{\K_{1,2} +
\A(t_r)}$ are plane waves. Their distortion by the
electron-electron and electron-core interactions appears in the
($\R_1, \R_2$)-dependent exponential phase factors in Eq.
(\ref{AmplB}).

Since $\Delta V_{shr}$ is a short-range potential and $\ket{g^+}$
is limited within a  characteristic ionic radius, the term
$\bra{\R_1 \R_2}\Delta V_{shr}\ket{g^+ \K}$  allows us to fix the
initial values of $\R_1$ and $\R_2$. The characteristic radius for
the partitioning of the Coulomb potential into the short-range and
long-range parts is set as  $r_0=r_{12}^{(0)} = 1/\left|
E_{g^+}\right|$. Therefore, we pull the exponential factor out of
the integral in Eq. (\ref{AmplB}) with $r_0=r_{12}^{(0)} =
1/\left| E_{g^+}\right|$ and $r_1 = r_2 =0$,
\begin{eqnarray}\label{AmplB2}
&& b(\K_1, \K_2, \K, t_r) \approx \bra{\K_1 + \A(t_r) \, \K_2 + \A(t_r)} \Delta V_{shr} \ket{g^+ \K}\exp\left[ -i\int_{t_r}^t \left\{ \frac 12 [\K_1 + \A(\tau)]^2 +\right.\right. \nonumber\\
&& \qquad  \left.\left. +\frac 12 [\K_2 + \A(\tau)]^2 +V_{ee,lng}(\R_{12}(\tau))+ V_{en,lng}(\R_{1}(\tau))+V_{en,lng}(\R_{2}(\tau)) \right\}d\tau\right].
\end{eqnarray}
Effects of the long-range tails of $V_{ee}$ and $V_{en}$ appear in
the exponent  while the collisional transition is govered by the
short-range interaction \cite{Smirnova_2008}. The states
$\ket{\K}$, so far, represent any convenient basis set of
continuum sates in the laser field.

To simplify the amplitude
$$
c(\K, t_r) = \bra{\K g^+} \U_s(t_r, t_i)\ket{gg},
$$
we note that the second electron is bound during the whole
evolution,  and hence we can simplify $c(\K, t_r)$ using single
active electron approximation. In this approximation, $\U_s(t_r,
t_i)$ describes one-electron dynamics in the self-consistent
potential of the ionic core,
$$
V_{sc} (\R_1) = \bra{g^+} V_{ee,lng}(\R_{12})+ V_{en,lng}(\R_{1})+V_{en,lng}(\R_{2})\ket{g^+}.
$$
The effective Hamiltonian for evolution between $t_i$ and $t_r$ is
$$
\h_{sc} (\R_1, t) = \hat{K}_1 + V_{sc}(\R_1) + V_L (\R_1, t),
$$
where $\hat{K}_1$ is the kinetic energy operator and $V_L(\R_1,
t)$ is  the interaction with the laser field. Now the amplitude
$c(\K, t_r)$ becomes
\begin{equation}\label{Ampl2}
c(\K, t_r) = -i \int_{t_i}^{t_r} dt_b \bra{\K} \U_{sc} (t_r, t_b) V_L(\R_1, t_b) \ket{g_D}\exp\left[i\left|E_{gg}\right| (t_b-t_i)\right],
\end{equation}
where $\ket{g_D} =\langle g^+_2 \ket{gg}$ is proportional to the Dyson orbital between the ground states of the neutral and ion.

The ionic potential contributes to the propagator in Eq. (\ref{Ampl2}) twice: when the electron leaves
the atom near $t_b$ and when it returns to the ionic core near $t_r$. The contribution ``on the way out''
introduces standard Coulomb correction
\cite{Perelomov_1966, Perelomov_1967_A, Perelomov_1967_B, Perelomov_1968, Popruzhenko2008a, Popruzhenko2008b}
to the ionization amplitude and hence affects the overall height of the final two-electron distribution.
The contribution of $V_{sc}$  ``on the way in'' affects the spatial structure of the recolliding wave packet.
As shown in Refs. \cite{Smirnova_2008, Smirnova_2007}, for short collision times the Coulomb-laser coupling
is small and $V_{sc}$ ``on the way in'' can be included in the adiabatic approximation,
\begin{eqnarray}\label{Ampl3}
c(\K, t_r)\ket{\K_{ev} g^+} \approx -i R_C \int_{t_i}^{t_r} dt_b \ket{\K_{ev} g^+}\bra{\K + \A(t_b) -\A(t_r)} V_L(t_b) \ket{g_D}\times \nonumber\\
 \exp\left[ -\frac i2 \int_{t_b}^{t_r} [\K + \A(\tau)-\A(t_r)]^2d\tau +i\left|E_{g^+}\right|(t_r - t_b) + i\left| E_{gg} \right|(t_b - t_i)\right]
\end{eqnarray}
Here $\ket{\K_{ev}}$ is the field-free continuum wave function in
the  eikonal approximation, which includes distortions of the
incoming plane wave with asymptotic momentum $\K$, $\bra{ \K +
\A(t_b) - \A(t_r)}$ is a plane wave, and $R_C$  is the Coulomb
correction to the ionization amplitude which compensates for
approximating $\bra{ \K + \A(t_b) - \A(t_r)}$ with a plane wave in
the matrix element $\bra{\K + \A(t_b) -\A(t_r)} V_L(t_b)
\ket{g_D}$.

Now, putting together Eq. (\ref{Ampl3}) and Eq. (\ref{AmplB2}) and
changing the integration variable $\K \to \K + \A(t_r)$, we arrive
at
\begin{eqnarray}\label{GenGamma}
&& a(\K_1, \K_2) \approx -\int_{t_i}^t dt_r \int_{t_i}^{t_r} dt_b \int d^3 \K \int d^3 \R_1 d^3 \R_2 \,   \nonumber\\
&& \times\exp\left[ -\frac i2 \int_{t_b}^{t_r} [\K + \A(\tau)]^2 d\tau
+i\left|E_{g^+}\right|(t_r - t_b) + i\left| E_{gg} \right|(t_b - t_i)-\right.\nonumber\\
&& \left. -i\int_{t_r}^t \left\{ \frac 12 [\K_1 + \A(\tau)]^2 + \frac 12 [\K_2 + \A(\tau)]^2 + V_{ee,lng}(\R_{12}(\tau))+
V_{en,lng}(\R_1( \tau))+ V_{en,lng}(\R_2( \tau)) \right\}d\tau\right] \nonumber\\
&&\times \bra{\K_1 + \A(t_r), \, \K_2 + \A(t_r)} \R_1 \R_2 \rangle\bra{\R_1 \R_2}
\Delta V_{shr} \ket{g^+,\, \K_{ev}+\A(t_r)}R_C\bra{\K+\A(t_b)} V_L(t_b) \ket{g_D}.
\end{eqnarray}
Note that if the Coulomb corrections $V_{ee,lng}$ and $V_{en,lng}$ are ignored in the exponent of Eq. (\ref{GenGamma}), then Eq. (\ref{GenGamma}) coincides with Eq. (\ref{SFA_S-matrix})  within exponential accuracy.

\section{The Correlated Two-electron Ionization within the SFA}\label{s3}

In this section, we find the correlated spectrum of the NSDI by
using the strong field approximation (SFA). In the next sections,
we will improve the SFA result by employing the SF-EVA \cite{Smirnova_2008}, i.e., the perturbation
theory in action with the SFA result as the zero-order
approximation.

Ignoring the Coulomb corrections in Eq. (\ref{GenGamma}) and performing the saddle-point calculations described in the Appendix,  we reach the usual SFA expression for the correlated NSDI spectra -- Eq. (\ref{SFASpectra}).

\begin{figure}
\begin{center}
\includegraphics[scale=0.6, angle=-90]{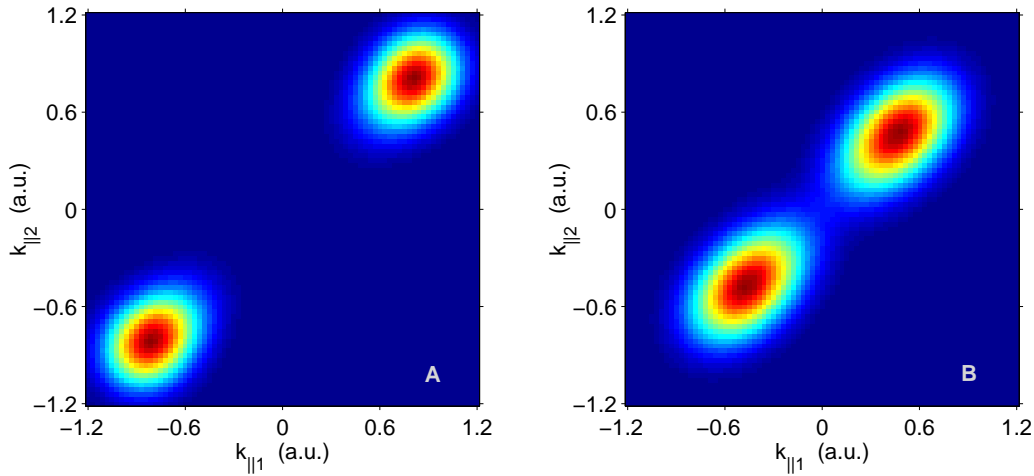}
\end{center}
\caption{Correlated two-electron spectra (\ref{SFASpectra}) of Ar (linear scale) within the
SFA at $7\times 10^{13}$ W/cm$^2$, 800 nm ($\k_1 = \k_2 = 0$)
(a) $\gamma=0$; (b) $\gamma = 1.373$. Maxima of figures correspond to probability densities: (a) $1.7\times10^{-6}$, (b) $2.9\times10^{-15}$.
}\label{Fig1}
\end{figure}

To illustrate the SFA results, we plot the two-electron correlated
spectrum for a system with $I_p$ of Ar in Fig. \ref{Fig1}. In Fig.
\ref{Fig1}.A we set $\gamma=0$. Such an SFA spectrum has a peak at
$\kk_1 = \kk_2 \approx -0.78$ a.u., which is the maximum of the
vector potential  $ -F/\omega \approx -0.78$ a.u. The last fact
has the following  interpretation: NSDI is most efficient when the
velocity of the incident electron is maximal. This is achieved
near the zero of the laser field, ${\bf E}(\varphi) = {\bf
F}\cos\varphi$, and the maximum of $\A(\varphi)$. An electron
liberated at this time could acquire the final drift velocity
$\approx -F/\omega$. However, including the correct value of the
Keldysh parameter $\gamma$ not only substantially shifts the peak
position (Fig. \ref{Fig1}.B), but also lowers the maximum by nine
orders of magnitude.

\section{Electron-Electron Interaction}\label{s4}

In this section, we demonstrate the changes in the correlated spectrum due to
 the electron-electron repulsion.

Coulomb corrections to the single-electron SFA theory were first
introduced by Perelomov, Popov, and Terent'ev
\cite{Perelomov_1967_B, Perelomov_1968} using the quasiclassical
(imaginary time) method (for reviews, see Refs. \cite{Popov2005,
Popov2004}). More recently, further improvements to this method
have been considered in Refs. \cite{Popruzhenko2008a,
Popruzhenko2008b}.  These improvements considered not only
subbarrier motion in imaginary time, but also the effects of the
Coulomb potential on the phase of the outgoing wave packet in
the classically allowed region. These improvements allowed the authors
of Refs. \cite{Popruzhenko2008a, Popruzhenko2008b} to obtain
quantitatively accurate results not only for ionization yields,
but also for the above threshold ionization spectra of direct
electrons (i.e., not including recollision). An alternative, but
conceptually similar, approach is the SF-EVA \cite{Smirnova_2008}.
Unlike the two previous methods, the SF-EVA allows a simple
treatment of the electron-electron and electron-ion interaction in
the two-electron continuum states.

 According to the SF-EVA, the contribution of the interaction potentials is calculated along the SFA trajectories,
$$
\R_{1,2}(\varphi) = \frac 1{\omega}\int_{\varphi_r^0}^{\varphi} \left[ \K_{1,2} + \A(\phi)\right] d\phi.
$$
Note that at the moment of recollision $\varphi_r^0$,  the
electrons are assumed to be at the origin, $ \R_{1,2}(\varphi_r^0)
={\bf 0}$. However, this does not cause any divergence since
according to Eq. (\ref{GenGamma}) we have to use the regularized
potential $V_{ee,lng}$.

From Eqs. (\ref{PotentialsIntro}) and (\ref{Potentials}), the
potential energy of electron-electron repulsion along these
trajectories is given by
\begin{eqnarray}\label{Vee2}
V_{ee,lng}(\varphi) &=&  \frac 1{r_{12}(\varphi)} \left(1- \exp\left[ - \frac{r_{12}(\varphi)}{r_{12}^{(0)}}\right] \right), \nonumber\\
r_{12}(\varphi)  &=& |\R_1(\varphi) - \R_2 (\varphi) |
 = \sqrt{ \left[ (\kk_1-\kk_2)\frac{\varphi-\varphi_r^0}{\omega} \right]^2 + \left[ (\k_1-\k_2)\frac{\varphi-\varphi_r^0}{\omega}
\right]^2}.
\end{eqnarray}
As discused in Sec. \ref{s1}, the parameter $r_{12}^{(0)}$ is set to the ionic radius, $r_{12}^{(0)}=1/I_p^{(2)}$.

The correlated spectrum, which accounts for the electro-electron interaction, has the form
\begin{eqnarray}\label{SpectrSFAEE}
\Gamma_{ee}(\K_1, \K_2) &\approx&  \exp\left(-\frac 2{\omega} \Im\left[S_{SFA}(\K_1, \K_2) + S_{ee}(\K_1, \K_2) \right]\right), \\
S_{ee}(\K_1, \K_2) &=&  \int_{\Re\varphi_r^0}^{\varphi_r^0} V_{ee,lng}(\varphi)d\varphi. \nonumber
\end{eqnarray}

\begin{figure}
\begin{center}
\includegraphics[scale=0.6]{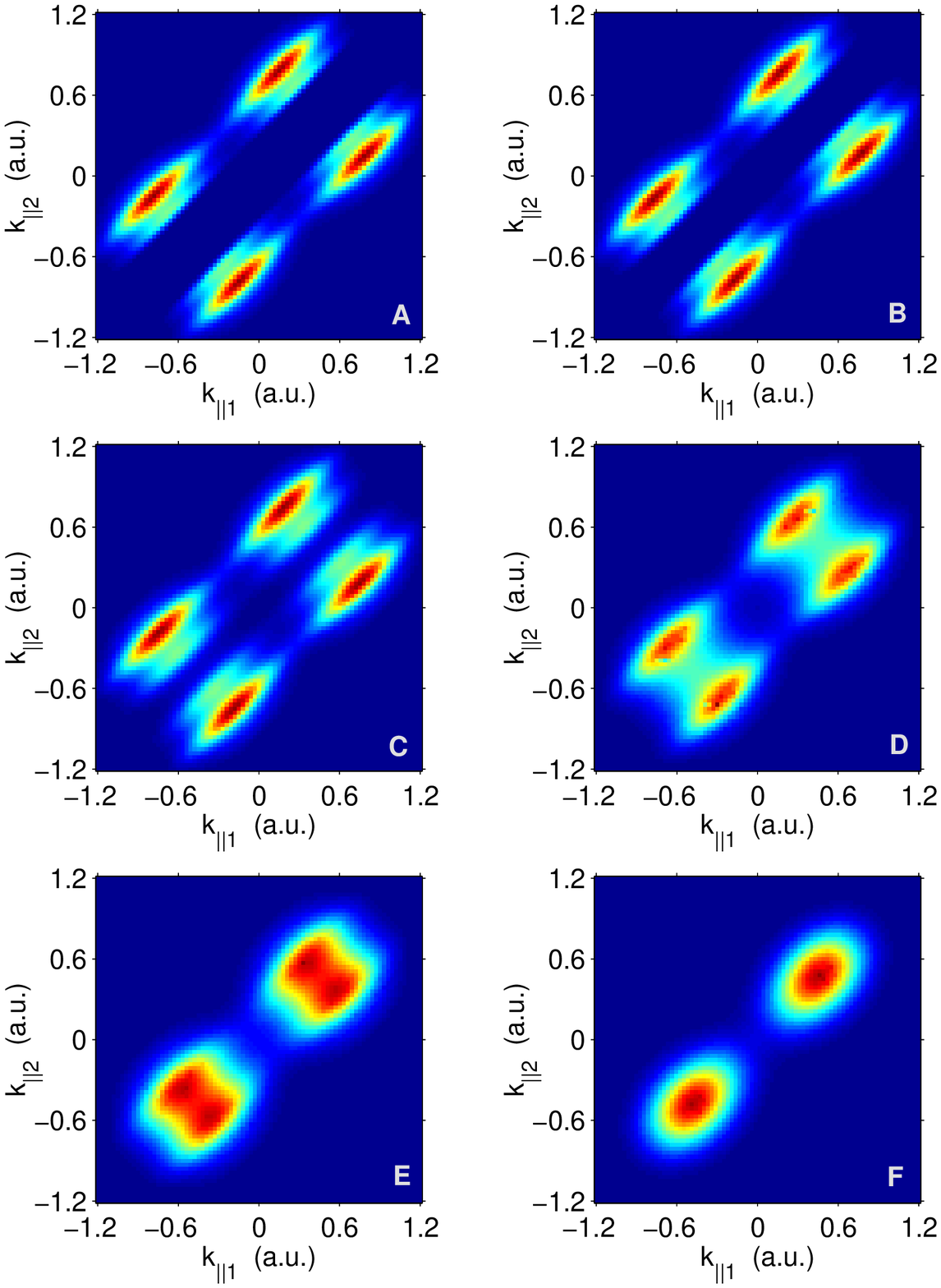}
\end{center}
\caption{ Role of electron-electron interaction. Correlated
spectra  of Ar (linear scale) at $7 \times 10^{13}$ W/cm$^2$, 800
nm are calculated using Eq. (\ref{SpectrSFAEE}) with $r_{12}^{(0)}
= 0.985$ a.u. ($\gamma=1.373$). Electron-core interaction is not
included.  Spectra are shown for different values of $\k$ (in
a.u.) for both electrons: (a) $\k_1 = \k_2 = 0$; (b) $\k_1 = 0,
\k_2 = 0.2$; (c) $\k_1 = -0.1, \k_2 = 0.2$; (d) $\k_1=-0.2, \k_2 =
0.3$; (e) $\k_1 = -0.3, \k_2 = 0.3$; (f) $\k_1 = -0.5, \k_2 =
0.5$. Maxima of figures correspond to probability densities: (a)
$9.1\times 10^{-19}$, (b) $6.5\times 10^{-19}$, (c)  $8.36\times
10^{-19}$, (d) $8.4\times 10^{-19}$, (e) $9.1\times 10^{-19}$, (f)
$2.8\times 10^{-20}$. } \label{Fig2}
\end{figure}

Figure \ref{Fig2} shows the contribution of electron-electron
repulsion to the spectra of NSDI for an atom with $I_p$ of
Ar (for experimental data see Ref. \cite{Liu_2008}).

Comparing Figs. \ref{Fig1} and \ref{Fig2}, we readily notice a
dramatic  influence of electron-electron interaction on the
correlated spectra. Electron-electron repulsion splits each SFA
peak into two peaks because, due to the Coulomb interaction, two
electrons cannot occupy the same volume. Note that the larger the
difference between the perpendicular momenta of both of the
electrons, the closer is the location of the peaks.

\section{Electron-Ion Interaction}\label{s5}

Now we include the electron-ion attraction. The potential energy of  electron-ion
interaction for the case of two electrons and a single core, after partitioning (\ref{PotentialsIntro}) and (\ref{Potentials}), is
\begin{eqnarray}\label{Vei2}
V_{en,lng}(\varphi) &=& -2\sum_{i=1}^2  \frac 1{r_{i}(\varphi)} \left(1- \exp\left[ - \frac{r_{i}(\varphi)}{r_0}\right] \right), \nonumber\\
r_{1,2}(\varphi) &=&
\sqrt{\left( \kk_{1,2}
\frac{\varphi-\varphi_r^0}{\omega} + \frac F{\omega^2}(
\cos\varphi - \cos\varphi_r^0 )\right)^2 + \left(
\k_{1,2}\frac{\varphi-\varphi_r^0}{\omega} \right)^2 }.
\end{eqnarray}
As far as the parameter $r_0$ is concerned, we set it equal to $r_{12}^{(0)}=1/I_p^{(2)}$.

Finally, the correlated spectrum of NSDI, which takes into account
both electron-electron and electron-ion interactions, reads
\begin{eqnarray}\label{SpectrSFAEEEI}
\Gamma_{ee+en}(\K_1, \K_2) &\approx&
\exp\left(-\frac 2{\omega} \Im\left[S_{SFA}(\K_1, \K_2) + S_{ee}(\K_1, \K_2) + S_{en}(\K_1, \K_2)\right]\right), \\
S_{en}(\K_1, \K_2) &=& \int_{\Re\varphi_r^0}^{\varphi_r^0} V_{en,lng}(\varphi)d\varphi. \nonumber
\end{eqnarray}

\begin{figure}
\begin{center}
\includegraphics[scale=0.6]{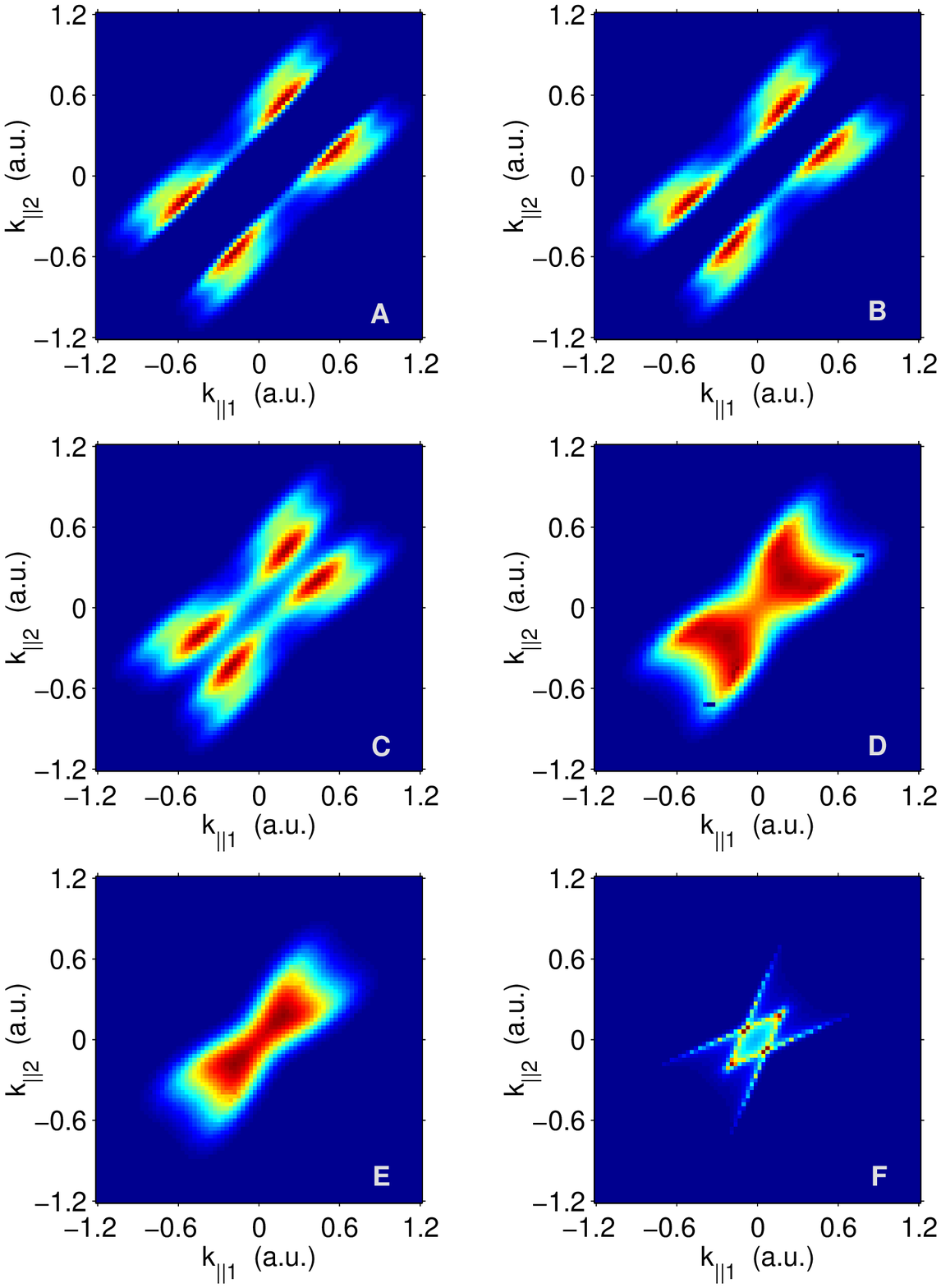}
\end{center}
\caption{ Role of electron-core interaction. Correlated spectra
of Ar (linear scale) at $7 \times 10^{13}$ W/cm$^2$, 800 nm are
calculated using Eq. (\ref{SpectrSFAEEEI}) with $r_{12}^{(0)}=r_0
= 0.985$ a.u. ($\gamma=1.373$). Both electron-elecron and
electron-core interactions are  included. Spectra are shown for
different values of $\k$ (in a.u.) for both electrons: (a) $\k_1 =
\k_2 = 0$; (b) $\k_1 = 0, \k_2 = 0.2$; (c) $\k_1 = -0.1, \k_2 =
0.2$; (d) $\k_1=-0.2, \k_2 = 0.3$; (e) $\k_1 = -0.3, \k_2 = 0.3$;
(f) $\k_1 = -0.5, \k_2 = 0.5$. Maxima of figures correspond to
probability densities: (a) $2.8\times10^{-5}$, (b)
$3.3\times10^{-5}$, (c) $6.1\times10^{-5}$, (d)
$1.5\times10^{-4}$, (e) $4.4\times10^{-4}$, (f)
$7.7\times10^{-4}$. } \label{Fig3}
\end{figure}

To illustrate the influence of the electron-ion attraction, we
have plotted the correlated spectra of Ar in Fig. \ref{Fig3} for
different perpendicular momenta. Comparing Figs. \ref{Fig2} and
\ref{Fig3}, we conclude that the  larger the difference between
the perpendicular momenta of the two electrons, the larger is the
contribution of the electron-ion interaction. Furthermore,
accounting for electron-ion attraction increases the probability of
NSDI by 15 orders of magnitude. This occurs because, as in
the case of single-electron ionization, electron-core interaction
significantly lowers an effective potential barrier. We can also
conclude that correlated spectra pictured in Figs. \ref{Fig3}.C,
\ref{Fig3}.D, and \ref{Fig3}.E have the biggest contribution to
the total probability of NSDI, which is an integral of the
probability density over momenta of both of the electrons. Note that,
on the one hand, the maximum of the probability density shown in
Fig. \ref{Fig3}.F is the largest among those presented in Fig.
\ref{Fig3}, and on the other hand, this maximum is localized in a few
pixels; therefore, the integral contribution of Fig. \ref{Fig3}.F
to the total probability is smaller than Fig. \ref{Fig3}.E.
Additionally, as one would expect, further increasing of $\k$ leads
to a decrease of probability density.  The correlated spectra in Fig.
\ref{Fig3}  agree  with the experimental data \cite{Liu_2008} in
quadrants one and three. The considered diagram (Fig. \ref{Fig4})
does not contribute to signals in quadrants two and four. Note
that taking into account a nonzero value of  $\gamma$ is vital to
achieve agreement with the experimental data.

From Eqs. (\ref{Vee2}) and (\ref{Vei2}), we can notice that  if
$r_{12}^{(0)} \to \infty$ and $r_0 \to \infty$, the Coulomb
corrections $V_{ee,lng}$ and $V_{en,lng}$ vanish, and the SFA
result is recovered. Therefore, we conclude that the radii
$r_{12}^{(0)}$ and $r_0$ contain the information about the initial
position of electrons after they emerged in the continuum.
Obviously, the intra-electron distance should be on the order of
an ion radius. 

\section{Conclusions}\label{s6}

The analytical quantum-mechanical theory
of NSDI within the deeply quantum regime, when the energy of the active electron driven by the
laser field is insufficient to collisionally ionize the parent
ion, has been formulated based on the SF-EVA approach. On the whole, the presented model
 agrees with available experimental data \cite{Liu_2008}. We have defined the
quantum-mechanical phase of birth of the active electron
(\ref{QuantPhaseBirth}), which accurately accounts for tunneling
of the recolliding electron in the regime where both the phases
$\varphi_r$ and $\varphi_b$ are complex.  Moreover, it has been
demonstrated that ignoring such a contribution of tunneling of the
active electron fails to agree with the experimental data.

Furthermore, our results show that any attempt to interpret NSDI
spectra  in this regime in terms of a simple SFA-based streaking
model would lead to wrong conclusions on the relative dynamics of
the two electrons.

The contributions of the electron-electron and electron-ion
interactions  have been analyzed. Both play an important and
distinct role in forming the shape of the correlated spectra.

The presented model is not able to give the correlated spectra in
quadrants two and four. It is because the considered process, when
two electrons detach {\it simultaneously} from the atom, does not
 contribute to that area. However, we incline to believe
that those parts of the correlated spectra are formed due to
recollision and excitation of the ion plus subsequent field
ionization \cite{Feuerstein_2001, Jesus_2004}, and it should be
noted  that this mechanism has been also observed in classical
simulations \cite{Haan2006}. We are planing to develop an
analytical model of such a mechanism in future papers.

\acknowledgments

We thank A. Rudenko, O. Smirnova, G.L. Yudin, and M.
Spanner  for highly stimulating discussions. Financial support to
D.I.B. by an NSERC SRO grant is gratefully acknowledged.

\appendix
\section{The phase of Ionization of the recolliding electron as a function of the phase of recollision}

Employing the SFA, we write the formula corresponding to the digram of NSDI (Fig. \ref{Fig4}) 
\begin{eqnarray}\label{SFA_S-matrix}
\ket{\Psi(t)} \sim \int_{t_i}^t dt_b \int_{t_b}^t dt_r \int d^3 \K \, \U(t, t_r) \frac 1{r_{12}} \ket{\K g^+}\bra{g^+ \K} \hat{V}_L(t_b) \ket{gg} \times \nonumber\\
\exp\left\{ -\frac i2 \int_{t_b}^{t_r} [\K + \A(\tau)]^2 d\tau +i|E_{g^+}|(t_r - t_b) + i|E_{gg}|(t_b-t_i) \right\},
\end{eqnarray}
where $\U(t, t_r)$ is the evolution operator of the studied system, $r_{12}$ is the distance between the electrons, $\hat{V}_L(t_b)$ is the interaction between the ionized electron and the laser field, and $E_{g^+}$ and $E_{gg}$ are energies of the states $\ket{g^+}$ and $\ket{gg}$, respectively.

We use the saddle point approximation (SPA) to calculate the
integrals over $\K$ and $t_b$ in Eq. (\ref{SFA_S-matrix}).  The phase
of the integral over $\K$ has the following form:
$$
S_1(\K) = -\frac 12 \int_{t_b}^{t_r}  \left[ \K + \A(\tau) \right]^2 d\tau.
$$
The saddle point of this integral is given by
\begin{equation}\label{Ksaddle}
\K^* = \frac{-1}{t_r - t_b} \int_{t_b}^{t_r} \A(\tau)d\tau,
\end{equation}
with the restriction $t_r \neq t_b$.  Note that generally
speaking, $\K^*$ can be complex since $t_b$,  as will be
clarified below, is complex for $\gamma \neq 0$.
 The phase of  the integral over $t_b$ in Eq. (\ref{SFA_S-matrix}) reads
$$
S_2 (t_b) = -\frac 12 \int_{t_b}^{t_r} [\K^* + \A(\tau)]^2 d\tau  +|E_{g^+}|(t_r - t_b) + |E_{gg}|(t_b-t_i).
$$
Hence, the saddle point $t_b(t_r)$ is a function of $t_r$ and given as a solution of the following equation
\begin{equation}\label{EqTbofTr}
\frac{\cos(\omega t_r) - \cos[\omega t_b(t_r)]}{\omega[t_r-t_b(t_r)]} + \sin[\omega t_b(t_r) ] = \pm i\gamma,
\end{equation}
where $\gamma$ is the Keldysh parameter for the first electron,
\begin{equation}\label{KeldyshParameter}
\gamma = \frac{\omega\sqrt{2(E_{g^+}-E_{gg})}}F.
\end{equation}
Methods of computing the saddle points [Eqs. (\ref{EqTbofTr}) and
(\ref{DeltaE})] have been  widely discussed in the literature (see,
for example, Refs. \cite{deMorissonFaria2005, deMorissonFaria2004,
deMorissonFaria2004B} and references therein). In the current
paper, we use a general and simple approach for identifying
correct saddle points between different solutions of the saddle-point equations in the complex plane (see Sec. \ref{s3}).

It is convenient to introduce the following phases: $\phi_b =
\omega t_b$ and $\varphi_r = \omega t_r$. According to Eq.
(\ref{EqTbofTr}),  the saddle point $\phi_b$ is a complex
double-valued function of $\varphi_r$ which can be given by
$\phi_b(+\gamma; \varphi_r)$ and $\phi_b(-\gamma; \varphi_r)$,
where $\gamma$ is the Keldysh parameter (\ref{KeldyshParameter}).
Here, the complex single-valued function $\phi_b(\gamma;
\varphi_r)$ is defined as a solution of the following
transcendental equation:
\begin{equation}\label{EqPhiBpfPhiR}
\frac{\cos \varphi_r - \cos \phi_b (\gamma; \varphi_r) }{\varphi_r - \phi_b(\gamma; \varphi_r)} + \sin \phi_b(\gamma; \varphi_r) = i\gamma.
\end{equation}
No analytical solution of such an equation is available.
The special case of the function $\phi_b(\gamma; \varphi_r)$ for $\gamma=0$,
\begin{equation}\label{EqPhiBpfPhiRG0}
\frac{\cos \varphi_r - \cos \varphi_b(\varphi_r) }{\varphi_r - \varphi_b(\varphi_r)} + \sin \varphi_b(\varphi_r) = 0,
\end{equation}
is very important because of the following two reasons.

First, the function $\varphi_b(\varphi_r)$ is a real valued
function for real $\varphi_r$ (and single-valued for any complex
$\varphi_r$); this allows one to interpret the motion of the first
electron in terms of classical trajectories. The
function $\varphi_b (\varphi_r)$ is defined on the interval $(\pi/2,
2\pi]$ because only during that interval can the free electron 
recollide with its parent ion. The function
$\varphi_b(\varphi_r)$ is bounded in the interval
$0\leqslant \varphi_b (\varphi_r)<\pi/2$. Second, the function $\varphi_b(\varphi_r)$ can be physically
understood as a tunneling limit ($\gamma \ll 1$) of
$\phi_b(\gamma; \varphi_r)$  (i.e., the low-frequency limit).

In terms of the laser phase, the
vector potential $\A(\varphi)$ is
$$
\A(\varphi) = - (\mathbf{F}/\omega) \sin\varphi.
$$

The difference between Eq. (\ref{EqPhiBpfPhiR}) and Eq.
(\ref{EqPhiBpfPhiRG0}),  which connect the phase of recollision
$\varphi_r$ with the phase of ionization $\varphi_b(\varphi_r)$ or
$\phi_b(\gamma; \varphi_r)$, is also important for the last step
 of NSDI -- the release of the  two electrons following the recollision at $\varphi_r$.

Calculating the integral over $t_r$ in Eq. (\ref{SFA_S-matrix}) by the SPA, we need to obtain the transition point
$\varphi_r^0$ for negligible $\gamma$. It is the solution of the  saddle-point equation
\begin{equation}\label{DeltaE}
\Delta E (\varphi_r^0) \equiv
\frac 12 \left[\K_1 + \A(\varphi_r^0)\right]^2 +  \frac 12 \left[\K_2 + \A(\varphi_r^0)\right]^2 - \frac 12 \left[\A(\varphi_r^0) - \A(\varphi_b( \varphi_r^0))\right]^2+I_p^{(2)}=0,
\end{equation}
such that
$$
\pi/2 < \Re \varphi_r^0 \leqslant 2\pi,
$$
where $I_p^{(2)} = |E_{g^+}|$ is the ionization potential of the second electron. For $\gamma \neq 0$, the equation is
\begin{equation}\label{DeltaEGamma}
\Delta E (\gamma; \varphi_r^0) \equiv
\frac 12 \left[\K_1 + \A(\varphi_r^0)\right]^2 +
\frac 12 \left[\K_2 + \A(\varphi_r^0)\right]^2 - \frac 12 \left[\A(\varphi_r^0) - \A(\phi_b(\gamma; \varphi_r^0))\right]^2+I_p^{(2)}=0,
\end{equation}
where $\phi_b(\gamma; \varphi_r^0)$ now depends on $\gamma$. Note that Eqs.
(\ref{DeltaE}) and (\ref{DeltaEGamma}) are basically Eq.
(\ref{EqT0}) written in slightly different notations.

If the solution of Eq. (\ref{DeltaE}) on the interval $(\pi/2,
2\pi]$ is real, then direct collisional ionization is possible.
However, we are interested in the deep quantum regime when the
following inequality is valid for the second electron:
$$
I_p^{(2)} > 3.17 U_p.
$$
By introducing the Keldysh parameter for the second electron
$
\gamma_2 = \sqrt{ \frac{I_p^{(2)}}{2U_p}},
$
we can write the last inequality as
\begin{equation}\label{ConditionCMI}
\gamma_2 > 1.26.
\end{equation}

Equation (\ref{ConditionCMI}) physically means that the returning
electron does not have enough energy to free the second electron.

When recollision energy is not sufficient for collisional
ionization, transition requires help from the laser field.
Mathematically, the arising integral is similar to those in the
Landau-Dykhne (LD) theory (see Sec. \ref{s0}). The energy gap $\Delta
E(\varphi_r)$ in Eqs. (\ref{DeltaE}) and (\ref{DeltaEGamma})
plays the role of the transition energy for the LD
transition [the term $E_i(t)-E_f(t)$ in Eq. (\ref{DykhneApp})].
The peculiarity of $\Delta E(\gamma; \varphi_r)$ given by Eq.
(\ref{DeltaEGamma}) is that it need not be  real even for real
$\varphi_r$, since in the term $\left[
A(\varphi_r)-A(\phi_b(\gamma; \varphi_r))\right]^2$ the phase
$\phi_b$ is complex. This subtle aspect underscores the important
difference between using the solutions of Eq. (\ref{EqPhiBpfPhiR})
or Eq. (\ref{EqPhiBpfPhiRG0}) for the phase of ionization
$\phi_b$. For classical trajectories, where $\varphi_b(\varphi_r)$
[Eq. (\ref{EqPhiBpfPhiRG0})] is real for real $\varphi_r$, $\Delta
E(\varphi_r)$ [Eq. (\ref{DeltaE})] is also real for real
$\varphi_r$. This is the standard assumption in the LD
theory. When the complex phase of ionization $\phi_b(\gamma;
\varphi_r)$ [Eq. (\ref{EqPhiBpfPhiR})] is used, i.e., when
``quantum'' trajectories for recollision are used, $\Delta
E(\gamma; \varphi_r)$ [Eq. (\ref{DeltaEGamma})]  need not be real
for real $\varphi_r$.  In \cite{Moyer_2001}, the
LD method has been generalized for this case, provided
that the complex function $\Delta E(\gamma; \varphi)$ satisfies
the Schwarz reflection principle (recently, this result was
confirmed and further generalized in Ref. \cite{Schilling_2006}).

The function $\Delta E(\varphi)$ (\ref{DeltaE}) obeys the Schwarz reflection
principle, i.e., $ \Delta E^*(\varphi^*) = \Delta E(\varphi).$
Hence, we can conclude that if $\varphi_r^0$ is the solution that
lies in the lower half-plane,  then $\left(\varphi_r^0\right)^*$ is
the solution that lies in the upper half-plane. Furthermore, it
can be easily proven that the following equation takes place for
any function $\Delta E(\varphi)$ which satisfies the Schwarz
reflection principle and any complex number $\varphi_r^0$:
$$
\Im\int_{\Re\varphi_r^0}^{\left(\varphi_r^0\right)^*} \Delta E(\varphi)d\varphi = -\Im\int_{\Re\varphi_r^0}^{\varphi_r^0}\Delta E(\varphi) d\varphi.
$$
From the previous equation, we can see that the transition points
that lie in the lower half-plane lead  to exponentially large
probabilities, which are unphysical.
Hereafter, let $\varphi_r^0$ denote  the  solution of the equation
$\Delta E(\varphi_r^0)=0$, which is the closest to the real axis
and lies in the upper-half plane.

Before continuing our discussion, let us point out the
following simple equalities, which follow from Eq.
(\ref{EqPhiBpfPhiR}):  $\Re[ \phi_b(+\gamma;
\varphi_r)]=\Re[\phi_b(-\gamma; \varphi_r)]$ and $\Im[
\phi_b(+\gamma; \varphi_r)]=-\Im[\phi_b(-\gamma; \varphi_r)]$ for
real $\varphi_r$. Furthermore, we obtain
\begin{equation}
 \label{GenSchwartPrin}
\phi_b^*(-\gamma; \varphi_r^*) = \phi_b(\gamma; \varphi_r), \quad
\Im\int_{\Re\varphi_r^0}^{\left(\varphi_r^0\right)^*} \Delta
E(\gamma;\varphi)d\varphi =
-\Im\int_{\Re\varphi_r^0}^{\varphi_r^0}\Delta E(-\gamma;\varphi)
d\varphi,
\end{equation}
where $E(\gamma;\varphi)$ is defined in Eq. (\ref{DeltaEGamma}).

Bearing in mind that formula (\ref{SFA_S-matrix}) must give an
exponentially small result (which implies that the transition
point must be located in the upper-half plane) and taking into
account Eq. (\ref{GenSchwartPrin}), we define
 the phase of ionization in the case of $\gamma\neq 0$ as
\begin{equation}\label{QuantPhaseBirth}
\Phi (\gamma; \varphi_r) = \left\{
\begin{array}{ll}
\phi_b(-\gamma; \varphi_r) & \mbox{if} \quad \Im\left( \varphi_r \right)>0, \\
\Re\left[\phi_b(\gamma;\varphi_r)\right]  &
\mbox{if}\quad \Im\left(\varphi_r\right)=0,\\
\phi_b(+\gamma; \varphi_r)  & \mbox{if} \quad \Im\left( \varphi_r \right)<0.
\end{array}
\right.
\end{equation}

Equation (\ref{QuantPhaseBirth}) is the most consistent definition of
the {\it quantum-mechanical} phase of ionization
 of the first electron
as a function of the phase of return. Generally speaking, there
was an ambiguity in selecting the value of $\Phi(\gamma;
\varphi_r)$ for real $\varphi_r$. However, we have chosen it in
such a way due to the following reason. The function
$\Im\left[\Phi(\gamma; \varphi_r)\right]$ has a jump discontinuity
on the real axis, but the function $\Re\left[\Phi(\gamma;
\varphi_r)\right]$ has a removable discontinuity that can be
removed by employing the equality
$$
\Re\left[\phi_b(\gamma;\varphi_r)\right] \equiv \frac 12 \left[\phi_b(\gamma; \varphi_r+i0) + \phi_b(-\gamma; \varphi_r-i0)\right]
\quad (\mbox{for real } \varphi_r ).
$$
Furthermore, the function $\Phi(\gamma; \varphi_r)$ obeys the
Schwarz reflection principle [$\Phi^*(\gamma; \varphi_r^*) =\Phi(\gamma; \varphi_r) $], and the following equality takes place:
$$
\Phi(0; \varphi_r) = \varphi_b( \varphi_r).
$$
It is essential that according to definition
(\ref{QuantPhaseBirth}), the function $\Phi(\gamma; \varphi_r)$ is
real-valued on the real axis and thus allows the identical
interpretation in terms of the classical trajectories as for
$\varphi_b( \varphi_r)$. Therefore, the
definition of $\varphi_r^0$ and inequality (\ref{ConditionCMI})
are unchanged in the case of $\gamma\neq  0$.

\bibliography{TEI}

\end{document}